\documentstyle[aps,prd]{revtex}
\input{epsf}

\begin{document}
\draft

\title{Perturbing Supersymmetric Black Hole}
\author{Hisashi Onozawa, Takashi Okamura, Takashi Mishima$ ^\dagger$, 
and
Hideki Ishihara}

\address{Department of Physics,
Tokyo Institute of Technology, \\
Oh-okayama, Meguro, Tokyo 152, Japan
\\
$ ^\dagger$Laboratories of Physics, College of Science and Technology, \\
Nihon University, Narashinodai, Funabashi, Chiba 274, Japan
}

\date{June 29, 1996}

\preprint{TIT-HEP-336/COSMO-74}

\maketitle

\begin{abstract}
An investigation of the perturbations of the
Reissner-Nordstr\"{o}m black hole in the $N=2$ supergravity is presented.
We prove in the extremal limit that the black hole responds
to the perturbation of each field in the same manner.
We conjecture that we can match the modes of the graviton,
gravitino and photon because of supersymmetry transformations.
\end{abstract}

\vspace{0.5cm}

The extreme Reissner-Nordstr\"{o}m black hole is very
important particularly in the context of  supergravities.
First, in supersymmetric theories the mass of the black hole is
bounded below by its charge \cite{GIB82}.
Then the extreme black hole saturates this bound.
This mass bound is surprisingly identical to the bound
imposed by the cosmic censorship conjecture.
Thus we can naturally avoid naked singularities in supergravity
theories \cite{KAL92b}.
At the same time, the extreme black hole has
the symmetry that the nonextreme black holes  do not have since
the spacetime admits the Killing spinor field,
which means the extreme black hole is invariant
under the supersymmetric transformations.
The state is a close analogy of a BPS saturated state for
supersymmetric particles.
Second, all orders of quantum corrections should vanish
in the extreme state.
This feature of the supergravity theories may lead to an insight to
the quantum effects around black holes \cite{KAL92a}.

So far, the extreme black holes in supergravities have been
discussed only in the case of static configurations.
However, to know more about the role of black holes in the supergravity
theories, we need to work on the dynamical aspects of black holes.
In the preceding paper \cite{ONO96},
we have investigated the quasinormal frequencies
of the extreme Reissner-Nordstr\"{o}m black hole in the
Einstein-Maxwell theory
and found an interesting fact that these frequencies were completely
identical for both the electromagnetic and the gravitational perturbation. 
The fact may be
something to do with the supersymmetry,
since the fields with different spins are related to each other.
Thus, we continue our analysis to the O(2) extended supergravity
\cite{FER76} to
investigate the perturbations of the Reissner-Nordstr\"{o}m black 
holes to get an insight on the black hole in
supergravity theories.

The O(2) extended supergravity field equations reduce to the usual
Einstein-Maxwell equations when the gravitino fields vanish.
By linearizing the field equations about a solution with vanishing
gravitino fields, we obtain a consistent set of equations
for the photon, gravitino and graviton fields on a background solution
of the Einstein-Maxwell equations.
When we consider the Reissner-Nordstr\"{o}m black hole
as a background solution,
these perturbation equations can all be reduced to the Regge-Wheeler type
equations.
For the graviton and photon \cite{CHA83}, the radial equations are
given by
\begin{equation}
  \left[ \frac{d^2}{dr^2_*}+ \omega^2 -V_s(r) \right] Z_s(r) =0,
  \label{eq:rw}
\end{equation}
where
\begin{eqnarray}
  \frac{dr}{dr_*} & = & \frac{\Delta}{r^2}, \\
  V_s    & = & \frac{\Delta}{r^5} \left[ A r 
                  -q_s +\frac{4Q^2}{r}\right], \\
  \Delta & = & r^2-2Mr+Q^2,\\
  q_1    & = & 3 M- \sqrt{9 M^2+4Q^2 (A-2)},\\
  q_2    & = & 3 M+ \sqrt{9 M^2+4Q^2 (A-2)},\\
  A      & = & j_s(j_s+1), \\
 & &           j_s=l+s,\hspace{1cm} (l=0,1,2,....).
\end{eqnarray}
The cases of 
$s=1$ and $s=2$ correspond to the electromagnetic
and the gravitational perturbation, respectively.
Here, $M$ and $Q$ are the mass and charge of the black hole,
and $j_s$ is an angular multipole index of the perturbation.
For the gravitino \cite{AIC81,TOR92}, the same equation as
Eq.(\ref{eq:rw}) is obtained.
On the other hand the potential is given by 
\begin{eqnarray}
  V_{\frac{3}{2}}      & = & G-\frac{dT_1}{dr_*}, \\
  G      & = & \frac{\Delta}{r^6} ( \lambda r^2 + 2 M r - 2 Q^2), \\
  T_1    & = & \frac{1}{F - 2Q}\left[\frac{dF}{dr_*} - \lambda
  \sqrt{\lambda^2 +1 }
       \right], \\
  F      & = & \frac{r^6}{\Delta^{1/2}}G, \\
 \lambda &=& (j_\frac{3}{2}-\frac{1}{2})(j_\frac{3}{2}+\frac{3}{2}), \\
 &&  j_{\frac{3}{2}}=l+\frac{3}{2}, \hspace{1cm} (l=0,1,2,...).
\end{eqnarray}

In the extremal limit under the normalization of $M=Q=1$, the
potentials for the three fields surprisingly 
reduce to similar forms:
\begin{eqnarray}
    V_1 &=& + (j_1+1) \frac{df}{dr_*}
   -4  f^3 + (j_1+1)^2 f^2,     \label{eq:v1}  \\
   V_{\frac{3}{2}} &=& + (j_\frac{3}{2} + \frac{1}{2})\frac{df}{dr_*}
                          -4  f^3 + (j_\frac{3}{2}+\frac{1}{2})^2 f^2, 
   \label{eq:v32} \\
   V_2 &=& - j_2 \frac{df}{dr_*}
         -4  f^3 + j_2^2 f^2, 
   \label{eq:v2}  \\
&&  j_s=l+s, \hspace{1cm} (l=0,1,2,...),
\end{eqnarray}
where $f$ is defined by
\begin{equation}
  f = \frac{r-1}{r^2},
\end{equation}
and 
\begin{equation}
  r_* = r-1 + \ln (r-1)^2 - \frac{1}{r-1}.
\end{equation}
These three potentials are related in the following way:
\begin{equation}
  V_1(r_*,j_1=j)=V_{\frac{3}{2}}(r_*,j_{\frac{3}{2}}=j+\frac{1}{2})=
                V_2(-r_*,j_2=j+1),
      \label{eq:mr}
\end{equation}
where $j$ is a positive integer.
The first equality is obvious in Eqs. (\ref{eq:v1}) and (\ref{eq:v32});
the potential of the Rarita-Schwinger field is identical to
that of the electromagnetic field if we shift a multipole index by 1/2.
The second equality is proved by using the readily verifiable relation
\begin{equation}
  f(r_*)=f(-r_*).
\end{equation}
Therefore $V_2$ can be obtained by reflecting $V_1$ or $V_{\frac{3}{2}}$
about $r_*=0$.
It is noteworthy that
this transformation corresponds to the exchange of the horizon
and infinity.
Eq. (\ref{eq:mr}) means that a scattering problem
for each perturbed field with a corresponding multiple index results
in the same transmission and reflection amplitudes.
Obviously,
the relation $V_1(r_*,j_1=j)=V_2(-r_*,j_2=j+1)$ proves our numerical results
in the previous paper \cite{ONO96}.

Next, we consider the cases of the nonextreme black holes.
It is necessary to see how the SUSY breaking of the black hole
influences on these three modes.
Since the potentials are complicated in the nonextreme cases,
it is difficult to analytically find a symmetry even if it exits.
Instead, we will numerically calculate the 
quasinormal frequencies, which are the resonant poles of the
problem.
The coincidence of these quasinormal modes is then 
a necessary condition for all fields to have the same
transmission amplitude.
The quasinormal modes of black holes have so far been calculated
through several methods 
\cite{CHA75a,LEA85,SCH85,IYE87a,NOL92,AND92b}.
We use the WKB method \cite{SCH85} to calculate the quasinormal frequencies.
The WKB frequencies are related to the potential in the following:
\begin{equation}
\omega_n^2 = {V_{0}} - i \left( n+\frac{1}{2}\right) \sqrt {-2V_{0}''},
\end{equation}
where the subscript 0 denotes the value at the top of the
potential and the subscript $n$ is 0 or a positive integer
called the mode number.
Thus we can evaluate how these potentials are close to each other.
Using the above formula, the first two quasinormal modes
of 
the electromagnetic, gravitino's and  gravitational perturbations
for $l=0,1,2$ are calculated for nearly extreme black holes.
As easily seen in Fig. 1, 
three modes approaches continuously as the charge of the black hole
increases and then meet in the extremal limit.
This is consistent with the fact that only the extreme black hole
preserves the supersymmetry.
Quasinormal frequencies of the electromagnetic
field move fastest in the
complex $\omega$ plane when we change the charge of the black hole,
because the electromagnetic field is most sensitive
to the change of the charge
and quasinormal frequencies of the gravitino field is
approximately the average value of the
electromagnetic and the gravitational frequencies.

The coincidence found among the graviton, gravitino and photon 
in the extremal limit can be considered as a
consequence of the supersymmetry.
It is well known that the extreme Reissner-Nordstr\"om black hole admits a
Killing spinor field \cite{GIB82}.
It means that the background solution is invariant under the supersymmetry
transformations with respect to the Killing spinor field.
This implies that all the perturbed fields can be related to
each other using
the supersymmetry transformations, which conserves the S-matrix.
Our results show that the gravitino is matched with the photon by adding
$1/2$ to the multipole index.
Since the multipole index $j_s$ here describes the total
angular momentum, 
this shift is explained by the supersymmetric transformation
that increases the spin by $1/2$ and hence increases the
total angular momentum by $1/2$.

Here we reach a new insight on the black hole in supergravities.
It is expected that 
for any supergravity theory the perturbed fields around the BPS
saturated black hole should have a similar behavior;
a scattering problem around the black hole is the same from one
field to another.
This is because all the fields can be matched with
each other using supersymmetry transformations and the character of
the black hole does not change whatever field we consider around it.
Recently the extreme black holes have been
intensively studied in the context of the strings and D-branes and
these studies have led to the first microscopic description
of the black hole entropy.
It is interesting to explicitly obtain the supersymmetric
transformations and relate them to Eq. (\ref{eq:mr}),
because our scenario can be clearly understood and will
shed new light on this subject.
This is under investigation.

\section*{Acknowledgments}

We would like to thank Professor G.W.~Gibbons for pointing out the
possible relation between our previous results and his joint
work with Professor C.M.~Hull.
We would like to thank Professor A.~Hosoya for his continuous encouragement
and Dr. B.~Meister for reading the manuscript.
The research is supported in part by the Scientific Research Fund of the
Ministry of Education.

\begin{figure}[htbp]
\begin{center}
  \leavevmode
  \epsfysize=7.5cm
  \epsfbox{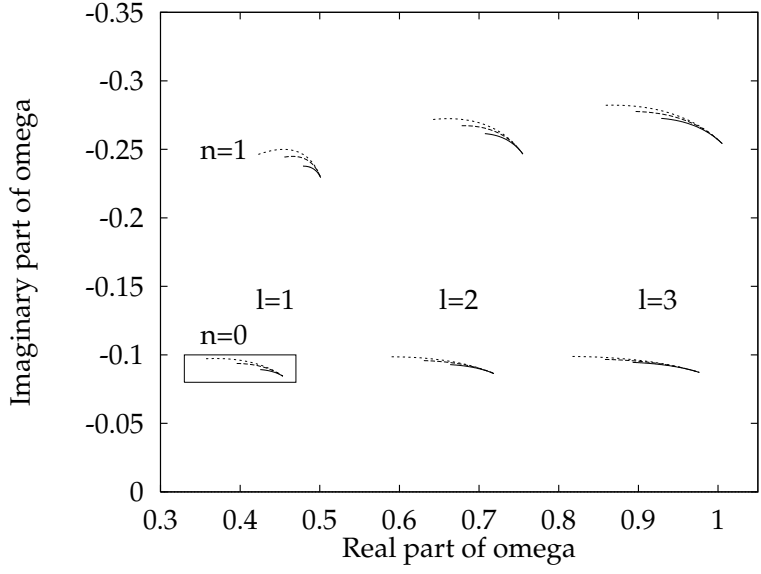}
\end{center}
\caption{
Solid lines, short dashed lines, and long dashed lines are the
trajectories of the first order WKB quasinormal frequencies
of the photon, gravitino, and graviton, respectively.
Each left endpoint of lines corresponds to the frequency
of a charged black hole of $Q=0.8$, and each right endpoint
corresponds to the frequency in the limit of maximal
charge.
A trajectory of each quasinormal frequency
meets at the right end with the two corresponding trajectories.
This means that supersymmety is recovered in the extremal limit.
}
\end{figure}

\begin{figure}[htbp]
\begin{center}
  \leavevmode
  \epsfysize=7.5cm
  \epsfbox{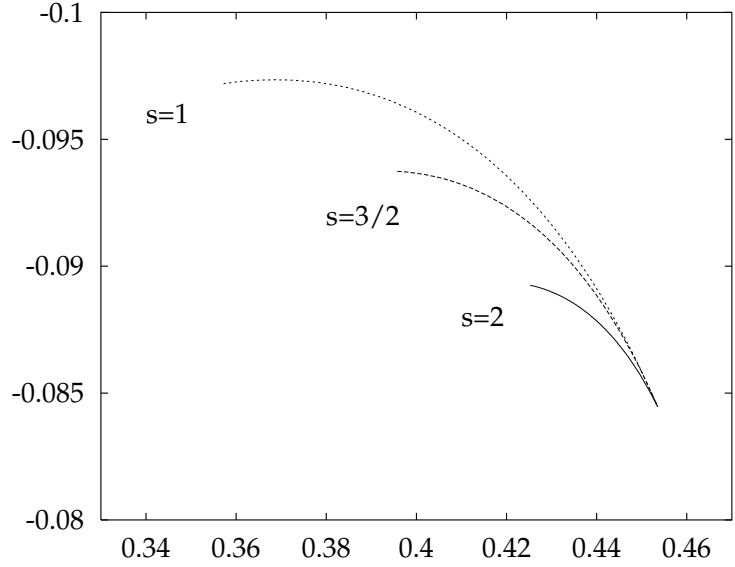}
\end{center}
\caption{
The bounded box in the above figure is zoomed in.
The quasinormal frequencies are equal only in the extremal limit.
This is consistent with the fact that the black hole preserves
supersymmetry only in the extremal limit.
}
\end{figure}

\end{document}